\documentclass[12pt]{article}
\begin{document}
\title{PLANCK SCALE TO COMPTON SCALE}
\author{B.G. Sidharth\\
International Institute for Applicable Mathematics \& Information Sciences\\
Hyderabad (India) \& Udine (Italy)\\
B.M. Birla Science Centre, Adarsh Nagar, Hyderabad - 500 063 (India)}
\date{}
\maketitle
\begin{abstract}
The Planck scale is considered to be a natural minimum scale, made up as it is solely of fundamental constants. However the Planck scale is well beyond the scales encountered in real life, these latter being at least of the order of the Compton scale. The Compton scale too is fundamental in the same sense -- it is made up of fundamental micro physical constants, though it is $10^{20}$ orders higher than the Planck scale. We discuss here a mechanism by which the Planck scale manifests itself at the real life Compton scale, this happening due to processes within the Planck time. Related issues are also discussed.
\end{abstract}
\section{Introduction}
Max Planck had noticed that, what we call the Planck scale today,
\begin{equation}
l_P = \left(\frac{\hbar G}{c^3}\right)^{\frac{1}{2}} \sim 10^{-33}cm\label{e1}
\end{equation}
is made up of the fundamental constants of nature and so, he suspected it played the role of a fundamental length. Indeed, modern Quantum Gravity approaches have invoked (\ref{e1}) in their quest for a reconciliation of gravitation with other fundamental interactions. In the process, the time honoured prescription of a differentiable spacetime has to be abandoned.\\
There is also another scale, made up of fundamental constants of nature, viz., the well known Compton scale,
\begin{equation}
l = e^2/m_ec^2 \sim 10^{-12}cm\label{e2}
\end{equation}
where $e$ is the electron charge and $m_e$ the electron mass.\\
The scale (\ref{e2}) has also played an important role in modern physics, though it is not considered as fundamental as the Planck scale. Nevertheless, the Compton scale (\ref{e2}) is close to reality in the sense of experiment, unlike (\ref{e1}), which is well beyond foreseeable experimental contact.\\
What we need to seek is a physical rationale for a route from (\ref{e1}) to (\ref{e2}).
\section{The Planck and Compton Scales}
We first give a rationale for the fact that the Planck scale would be a minimum scale in the universe. Our starting point \cite{bgsijmpe} is the model for the underpinning at the Planck scale for the universe. This is a collection of $N$ Planck scale oscillators (Cf.refs.\cite{psu,psp,uof,gip,ng} for details). We do not need to specfify $N$. We have in this case the following well known relations
$$R = \sqrt{N}l, Kl^2 = kT,$$
\begin{equation}
\omega^2_{max} = \frac{K}{m} = \frac{kT}{ml^2}\label{e3}
\end{equation}
In (\ref{e3}), $R$ is of the order of the diameter of the universe, $K$ is the analogue of the spring constant, $T$ is the effective temperature while $l$ is the analogue of the Planck length, $m$ the analogue of the Planck mass and $\omega_{max}$ is the frequency--the reason for the subscript $max$ will be seen below. We do not yet give $l$ and $m$ their usual values as given in (\ref{e1}) for example, but rather try to deduce these values.\\
We now use the well known result that the individual minimal oscillators are black holes or mini universes as shown by Rosen \cite{rosen}. So using the well known Beckenstein temperature formula for these primordial black holes \cite{ruffini}, that is
$$kT = \frac{\hbar c^3}{8\pi Gm}$$
in (\ref{e3}) we get,
\begin{equation}
Gm^2 \sim \hbar c\label{e4}
\end{equation}
which is another form of (\ref{e1}). We can easily verify that (\ref{e4}) leads to the value $m \sim 10^{-5}gms$. In deducing (\ref{e4}) we have used the typical expressions for the frequency as the inverse of the time - the Compton time in this case and similarly the expression for the Compton length. However it must be reiterated that no specific values for $l$ or $m$ were considered in the deduction of (\ref{e4}).\\
We now make two interesting comments. Cercignani and co-workers have shown \cite{cer1,cer2} that when the gravitational energy becomes of the order of the electromagnetic energy in the case of the Zero Point oscillators, that is
\begin{equation}
\frac{G\hbar^2 \omega^3}{c^5} \sim \hbar \omega\label{e5}
\end{equation}
then this defines a threshold frequency $\omega_{max}$ above in which the oscillations become chaotic.\\
Secondly from the parallel but unrelated theory of phonons \cite{huang,rief}, which are also bosonic oscillators, we deduce a maximal frequency given by
\begin{equation}
\omega^2_{max} = \frac{c^2}{l^2}\label{e6}
\end{equation}
In (\ref{e6}) $c$ is, in the particular case of phonons, the velocity of propagation, that is the velocity of sound, whereas in our case this velocity is that of light. Frequencies greater than $\omega_{max}$ in (\ref{e6}) are meaningless.  We can easily verify that (\ref{e5}) and (\ref{e6}) give back (\ref{e4}).\\
Finally we can see from (\ref{e3}) that, given the value of $l_P$ and using the value of the radius of the universe, viz., $R \sim 10^{27}$, we can deduce that, 
\begin{equation}
N \sim 10^{120}\label{e7}
\end{equation}
In a sense the relation (\ref{e4}) can be interpreted in a slightly different vein as representing the scale at which all energy- gravitational and electromagnetic becomes one.\\
The Compton scale comes as a Quantum Mechanical effect, within which we have zitterbewegung effects and a breakdown of Causal Physics \cite{diracpqm}. Indeed Dirac had noted this aspect in connection with two difficulties with his electron equation. Firstly the speed of the electron turns out to be the velocity of light. Secondly the position coordinates become complex or non Hermitian. His explanation was that in Quantum Theory we cannot go down to arbitrarily small spacetime intervals, for the Heisenberg Uncertainty Principle would then imply arbitrarily large momenta and energies. So Quantum Mechanical measurements are an average over intervals of the order of the Compton scale. Once this is done, we recover meaningful physics. All this has been studied afresh by the author more recently, in the context of a non differentiable spacetime and noncommutative geometry.\\
Weinberg also notices the non physical aspect of the Compton scale \cite{weinberg}. Starting with the usual light cone of Special Relativity and the inversion of the time order of events, he goes on to add, and we quote at a little length and comment upon it, ``Although the relativity of temporal order raises no problems for classical physics, it plays a profound role in quantum theories. The uncertainty principle tells us that when we specify that a particle is at position $x_1$ at time $t_1$, we cannot also define its velocity precisely. In consequence there is a certain chance of a particle getting from $x_1$ to $x_2$ even if $x_1 - x_2$ is spacelike, that is, $| x_1 - x_2 | > |x_1^0 - x_2^0|$. To be more precise, the probability of a particle reaching $x_2$ if it starts at $x_1$ is nonnegligible as long as
$$(x_1 - x_2)^2 - (x_1^0 - x_2^0)^2 \leq \frac{\hbar^2}{m^2}$$
where $\hbar$ is Planck's constant (divided by $2\pi$) and $m$ is the particle mass. (Such space-time intervals are very small even for elementary particle masses; for instance, if $m$ is the mass of a proton then $\hbar /m = 2 \times 10^{-14}cm$ or in time units $6 \times 10^{-25}sec$. Recall that in our units $1 sec = 3 \times 10^{10}cm$.) We are thus faced again with our paradox; if one observer sees a particle emitted at $x_1$, and absorbed at $x_2$, and if $(x_1 - x_2)^2 - (x_1^0 - x_2^0)^2$ is positive (but less than or $=\hbar^2 /m^2)$, then a second observer may see the particle absorbed at $x_2$ at a time $t_2$ before the time $t_1$ it is emitted at $x_1$.\\
``There is only one known way out of this paradox. The second observer must see a particle emitted at $x_2$ and absorbed at $x_1$. But in general the particle seen by the second observer will then necessarily be different from that seen by the first. For instance, if the first observer sees a proton turn into a neutron and a positive pi-meson at $x_1$ and then sees the pi-meson and some other neutron turn into a proton at $x_2$, then the second observer must see the neutron at $x_2$ turn into a proton and a particle of negative charge, which is then absorbed by a proton at $x_1$ that turns into a neutron. Since mass is a Lorentz invariant, the mass of the negative particle seen by the second observer will be equal to that of the positive pi-meson seen by the first observer. There is such a particle, called a negative pi-meson, and it does indeed have the same mass as the positive pi-meson. This reasoning leads us to the conclusion that for every type of charged particle there is an oppositely charged particle of equal mass, called its antiparticle. Note that this conclousion does not obtain in nonrelativistic quantum mechanics or in relativistic classical mechanics; it is only in relativistic quantum mechanics that antiparticles are a necessity. And it is the existence of antiparticles that leads to the characteristic feature of relativistic quantum dynamics, that given enough energy we can create arbitrary numbers of particles and their antiparticles.''\\
We note however that there is a nuance here which distinguishes Weinberg's explanation from that of Dirac. In Weinberg's analysis, one observer sees only protons at $x_1$ and $x_2$, whereas the other observer sees only neutrons at $x_1$ and $x_2$ while in between, the first observer sees a positively charged pion and the second observer a negatively charged pion. We will come back to this point later but remark that Weinberg's explanation is in the spirit of the Feynman-Stuckleberg diagrams.
\section{The Transition}
We now address the question of the mechanism by which there is a transition from the Planck scale to the Compton scale. For this we will need a relation that is well known in the literature \cite{ijmpa,ijtp}
\begin{equation}
G = \Theta/t\label{ex}
\end{equation}   
It may be mentioned that this relation (\ref{ex}) shows the gravitational constant as varying with time. This dependence also features in the Dirac Cosmology \cite{narlikar}.\\
We now observe the following: It is known that for a Planck mass $m_P \sim 10^{-5}gm$, all the energy is gravitational and in fact we have, as in (\ref{e4}),
$$Gm^2_P \sim e^2$$
For such a mass the Schwarzschild radius is the Planck length or Compton length for a Planck mass 
\begin{equation}
\frac{Gm_P}{c^2} = l_P \sim \hbar/m_P c \sim 10^{-33}cm\label{e8}
\end{equation}
We can compare (\ref{e8}) with (\ref{e2}) which defines $l$ as what may be called the ``electromagnetic Schwarzschild'' radius viz., the Compton wavelength, when $e^2$ is seen as an analogue of $Gm^2$. To push these considerations further, we have from the theory of black hole thermodynamics [11, 12] for any arbitrary mass $m$, the well known Beckenstein temperature given by
\begin{equation}
T = \frac{\hbar c^3}{8\pi km G}\label{ea6}
\end{equation}
Equation (\ref{ea6}) gives the thermodynamic temperature of a Planck mass black hole. Further, in this theory as is known \cite{ruffini},
\begin{equation}
\frac{dm}{dt} = - \frac{\beta}{m^2},\label{e9}
\end{equation}
where $\beta$ is given by
$$\beta = \frac{\hbar c^4}{(30.8)^3 \pi G^2}$$
This leads back to the usual black hole life time given by
\begin{equation}
t = \frac{1}{3\beta} m^3 = 8.4 \times 10^{-24} m^3 secs\label{ea7}
\end{equation}
Let us now factor in the time variation (\ref{ex}) of $G$ into (\ref{e9}). Equation (\ref{e9}) now becomes
$$m^2 dm = -B \, \mu^{-2} t^2 dt, B \equiv \frac{\hbar c^4}{\lambda^3 \pi}, \,  \mu \equiv \frac{l c^2 \tau}{m}, \lambda^3 = (30.8)^3 \pi$$
Whence on integration we get
\begin{equation}
m = \frac{\hbar}{\lambda \pi^{1/3}} \left\{ \frac{1}{l^6}\right\}^{1/3} t \, = \frac{\hbar}{\lambda \pi^{1/3}} \frac{1}{l^2} t\label{e10}
\end{equation}
If we use the pion Compton time for $t$, in (\ref{e10}), we get for $m$, the pion mass. In other words, due to (\ref{ex}), the evanescent Planck mass decays into a stable elementary particle. We will return to this conclusion from an alternative viewpoint.
\section{Characterizing the Compton Scale}
As we have seen, within the Compton scale, an observer can percieve a proton going from $A$ to $B$, while another observer could describe the same event as a neutron going from $B$ to $A$. The first observer, in this particular example also sees a positive pion, whereas the second observer sees a negative pion. The question that arises is, what has happened to the charge? The only conserved quantity would be the baryon number. We will analyse this point now but remark that there is already an answer (Cf.ref.\cite{cu}). The electromagnetic nature of interaction is what we percieve outside the Compton wavelength. Within the Compton scale, the interaction shows up as the interquark QCD potential, conserving the baryon number.\\
The point is, that once it is recognized that the Compton wavelength is a cut off, as discussed in detail (Cf.ref.\cite{uof}), this implies a noncommutative geometry
\begin{equation}
[dx^\mu , dx^\nu ] \approx \beta^{\mu \nu} l^2 \ne 0\label{e14}
\end{equation}
While Equation (\ref{e14}) is true for any minimum cut off $l$ as shown by Snyder nearly sixty years ago, it is most interesting and leads to physically meaningful relations, when $l$ is at the Compton scale. In any case given (\ref{e14}), the usual invariant line element, 
\begin{equation}
ds^2 = g_{\mu \nu} dx^\mu dx^\nu\label{e15}
\end{equation}
has to be written in terms of the symmetric and nonsymmetric combinations for the product of the coordinate differentials. That is the right side of Equation (\ref{e15}) would become
$$\frac{1}{2} g_{\mu \nu} \left[\left(dx^\mu dx^\nu + dx^\nu dx^\mu\right) + \left(dx^\mu dx^\nu - dx^\nu dx^\mu\right)\right],$$
In effect we would have
\begin{equation}
g_{\mu \nu} = \eta_{\mu \nu} + kh_{\mu \nu}\label{e16}
\end{equation}
So the noncommutative geometry introduces an extra term, that is the second term on the right side of (\ref{e16}), which plays the role of the usual energy momentum tensor. All this ofcourse is at the Compton scale of an elementary particle. As in the case of General Relativity \cite{ohan}, but this time rememebring that neither the coordinates nor the derivatives commute we have
$$\partial_\lambda \partial^\lambda h^{\mu \nu} - (\partial_\lambda \partial^\nu h^{\mu \lambda} + \partial_\lambda \partial^\mu h^{\nu \lambda})$$
\begin{equation}
-\eta^{\mu \nu} \partial_\lambda \partial^\lambda h + \eta^{\mu \nu}\partial_\lambda \partial_\sigma h^{\lambda \sigma} = - k T^{\mu \nu}\label{e17}
\end{equation}
It must be stressed that the non commutativity of the space coordinates has thrown up the analogue of the energy momentum tensor of General Relativity, viz., $T^{\mu \nu}$. We identify this with the energy momentum tensor.\\
Remembering that $h_{\mu \nu}$ is a small effect, we can use the methods of linearized General Relativity \cite{ohan}, to get from (\ref{e17}), 
\begin{equation}
g_{\mu v} = \eta_{\mu v} + h_{\mu v}, h_{\mu v} = \int
\frac{4T_{\mu v}(t-|\vec x - \vec x'|,\vec x')}{|\vec x - \vec x'|}
d^3x'\label{e18}
\end{equation}
It was shown several years ago that for distances $|\vec{x} - \vec{x'}|$ much greater than the distance $\vec{x'}$, that is well outside the Compton wavelength, we can recover from (\ref{e18}) the electromagnetic potential.\\
Let us now see what happens when $|\vec x| \sim |\vec x'|.$ In this case, we have from (\ref{e18}), expanding in a Taylor series about $t$,
\begin{eqnarray}
h_{\mu v} = 4 \int \frac{T_{\mu v}(t,\vec x')}{|\vec x - \vec x'|}d^3 x'+
(\mbox{terms   independent  of}\vec x) + 2 \nonumber \\
\int \frac{d^2}{dt^2} T_{\mu v} (t,\vec x'). |\vec x - \vec x'| d^3 x' +
0(|\vec x - \vec x'|^2)\label{e19}
\end{eqnarray}
The first term gives a Coulombic $\frac{\alpha}{r}$ type interaction except
that the coefficient $\alpha$ is of much greater magnitude as compared to
the gravitational or electromagnetic case, because in this approximation,
in an expansion of $(1/|\vec x - \vec x'|),$ all terms are of comparable order. The second term on the right side of (\ref{e19}) is of no dynamical value as it is independent of $\vec{x}$. The third term however is of the form constant $\times r$. That is the potential (\ref{e19}) is exactly of the form of the QCD potential \cite{lee}
\begin{equation}
-\frac{\alpha}{r} + \beta r\label{e20}
\end{equation}
In (\ref{e20}) $\alpha$ is of the order of the mass of the particle as follows from (\ref{e19}) and the fact that $T^{\mu \nu}$ is the energy momentum tensor given by 
\begin{equation}
T^{\mu \nu} = \rho u^\mu u''\label{e21}
\end{equation}
where $u^\mu$ represented the four velocity. Remembering that we are within a sphere of radius given by the Compton length where the velocities equal that of light, we have equations
\begin{equation}
|\frac{du_v}{dt}| = |u_v|\omega\label{e22}
\end{equation}
\begin{equation}
\omega = \frac{|u_v|}{R} = \frac{2mc^2}{\hbar}\label{e23}
\end{equation}
It is interesting to note that we get (\ref{e22}) in the theory of the Dirac equation \cite{diracpqm}, viz., 
$$\imath \hbar \frac{d}{dt} (c\alpha_\imath) = -2mc^2 (c\alpha_\imath),$$
Using (\ref{e21}), (\ref{e22}) and (\ref{e23}) we get
\begin{equation}
\frac{d^2}{dt^2} T^{\mu v} = 4 \rho u^\mu u^v \omega^2 = 4 \omega^2 T^{\mu v}\label{e24}
\end{equation}
Equation (\ref{e24}) too is observed in the Dirac theory, which is an alternative derivation. Whence, as can be easily verified, $\alpha$ and $\beta$ in (\ref{e20}) have the correct values required for the QCD potential (Cf. also \cite{cu}). Alternatively $\beta r$ can be obtained, as in the usual theory by a comparison with the Regge angular momentum mass relation: It is in fact the constant string tension like potential which gives quark confinement and its value is as in the usual theory.\\
In other words (\ref{e20}) now gives the interquark potential with the Coulombic and confining terms and the right values for the coupling constants..\\
It can now be seen why within the Compton scale, the charge disappears--at this scale we deal with quarks, with a conservation of the baryon number. This would explain the apparently paradoxical feature of Weinberg's interpretation.\\
\section{Discussion}
1. The purport of Equation (\ref{e10}) is that an initial Planck oscillator (with Planck mass) decays in a time $\tau$, the Compton time of an elementary particle into an elementary particle which is stable in comparison to the Planck mass. This conclusion can be obtained from an alternative point of view. Let us consider the initial Planck scale oscillator as a Gaussian wave packet. Then it is well known (Cf.ref.\cite{powell}) that the wave packet spreads with time, and at time $t$, its width is given by
\begin{equation}
\Delta x = \frac{\sigma}{\sqrt{2}} \sqrt{1 + \frac{\hbar^2 t^2}{\sigma^4 m^2}}\label{e25}
\end{equation}
where $\sigma$ is $\sim \Delta x$. This shows that after a time $t$ which is of the order of the Compton time of a particle of mass $m$, the width of the original packet decays into a width which is of the order of the Compton scale of the particle of mass $m$ which is the result in (\ref{e10}).\\
2. We now consider the above results (\ref{e10}) and (\ref{e25}) in the following context: It has been shown by the author that the photon could be considered to have a mass $\sim 10^{-65}gms$ and Compton wavelength $\sim 10^{28}cms$, that is the radius of the universe (Cf.ref.\cite{uof,bhtd}). It is interesting that the Planck oscillator can now be seen to decay via (\ref{e10}) or (\ref{e25}) to the photon mass or equivalently the energy $10^{-33}eV$. It is quite remarkable that this cosmic footprint has indeed been found rather recently \cite{ijmp,newphys}. This is the observational confirmation of the above ideas.\\
3. It may be mentioned that as shown elsewhere (Cf.ref.\cite{uof}) in the successful Planck oscillator model referred to above (Cf. also ref.\cite{pk}), while the collection of $10^{120}$ Planck oscillators is the highest excited state, $10^{80}$ elementary particle scale is a lower energy stable state and $10^{120}$ photon scale is the lowest energy and stablest state.\\
4. From a geometrical point of view, the Weinberg characterization of the Compton scale could be expressed in the following way: Special Relativity gives the hyperbolic geometry of the Minkowski metric. With the introduction of the Quantum Mechanical Compton wavelength effect as in Section 2, at the Compton time scale, the space geometry itself becomes hyperbolic, because effectively there is a new coordinate $\vec{r}'$ such that 
$$\vec{r}^2 - \vec{r}'^2 = \tau^2$$
where
$$\vec{r}'^2 < \sim = l^2$$
with a simple transformation of the space coordinates, for a simplicity in one dimension, this can be written in the form
$$x \bar{x} \approx l^2$$
The Compton wavelength effect is seen to be identical to that of a non commutative geometry with $\bar{x}$ playing the role of a momentum with a suitable dimensional constant viz.,
$$\bar{x} = \frac{l^2}{\hbar} p_x$$
This is exactly the result obtained from a different route \cite{ffp4}.\\
Looking at it another way the usual Minkowski geometry defines the time like region and the space like region separated by the boundary of the light cone. Events in the time like region are causally connected and retain their temporal order whereas events falling outside the light cone in the space like region can have their temporal order reversed. With the introduction of the Quantum Mechanical Compton wavelength effect effectively some of the space like region now becomes a part of the time like region showing apparently super luminal effects within the Compton region.\\
5. It may be mentioned that the considerations in Section 4 lead via a non commutative geometry to an energy momentum tensor in (\ref{e17}). We can take this to be the origin of mass and spin itself, for we have
$$m = \int T^{00} d^3 x$$
and via
$$S_k = \int \epsilon_{klm} x^l T^{m0} d^3 x$$
the equation
$$S_k = c < x^l > \int \rho d^3 x$$
while $m$ above can be immediately identified with the mass, the last equation gives the Quantum Mechanical spin if we remember that
$$\langle x^l \rangle = \frac{\hbar}{2mc}$$
6. It is interesting to apply the Compton wavelength considerations above to the photon itself with the mass suggested above and the consequent Compton length, viz., the radius of the universe itself. This would then lead to the following scenario: An observer would see a photon leaving a particle $A$ and then reaching another planet $B$, while a different observer would see exactly the opposite. The distinction between the advanced and retarded potentials of the old electromagnetic theory thus gets mixed up. In fact there is an immediate explanation for this in the Instantaneous Action At a Distance theory first discussed by Feynman and Wheeler and more recently by Hoyle and Narlikar \cite{hoyle}. In this case the usual causal electromagnetic field would be given by half the sum of the advanced and retarded fields, indeed as noted by Dirac many years ago.  

\end{document}